\documentclass[12pt,preprint]{aastex}

\def\gs{\mathrel{\raise0.35ex\hbox{$\scriptstyle >$}\kern-0.6em
\lower0.40ex\hbox{{$\scriptstyle \sim$}}}}
\def\ls{\mathrel{\raise0.35ex\hbox{$\scriptstyle <$}\kern-0.6em
\lower0.40ex\hbox{{$\scriptstyle \sim$}}}}

\shortauthors{Owen \& Morrison}
\shorttitle{Deep SWIRE Field I. 20cm Continuum}

\begin{document}

\title{The Deep SWIRE Field}
\title{I. 20cm Continuum Radio Observations: A Crowded Sky}

\author{
Frazer\,N.\ Owen,\altaffilmark{1}, 
\& G.\,E.\ Morrison,\altaffilmark{2,3}}
\altaffiltext{1}{National Radio Astronomy Observatory, P.\ O.\ Box O,
Socorro, NM 87801 USA.; fowen@nrao.edu The National Radio Astronomy
Observatory is facility of the National Science Foundation operated
under cooperative agreement by Associated Universities Inc.}
\altaffiltext{2}{Institute for Astronomy, University of Hawaii, 96822,
  USA}
\altaffiltext{3}{Canada-France-Hawaii Telescope Corporation, Kamuela,
  Hawaii, 96743, USA}

\setcounter{footnote}{4}

\begin{abstract}

	We present results from deep radio observations taken with 
the VLA at a center frequency of 1400 MHz covering a region of the
SWIRE Spitzer Legacy survey, centered
at 10$^h$46$^m$00$^s$, 59\arcdeg01\arcmin00\arcsec (J2000). The reduction
and cataloging of radio
sources are described. An electronic catalog of the sources detected
above 5 sigma is also presented. 

The survey presented is the deepest so far in terms of the radio source
density on the sky. Perhaps surprisingly, the sources down to the bottom
of the catalog appear to have median angular sizes still $> 1$ arcsecond, like
their cousins 10-100 times stronger. The shape of the differential 
log N - log S counts also seems to require a correction for the finite
sizes of the sources to be self-consistent. If the log N - log S
normalization remains constant
at the lowest flux densities, there are about 6 sources per square arcminute
at 15 $\mu$Jy at 20cm. Given the finite source size this implies that we
may reach the natural confusion limit near 1 $\mu$Jy. 

\end{abstract}

\keywords{cosmology: observations ---  galaxies:
evolution --- galaxies: starburst --- infrared: galaxies
galaxies}

\section{Introduction}

     Understanding the growth and evolution of galaxies is
one of the key goals of modern astronomy. A new generation
of multi-wavelength surveys from X-ray to radio wavelengths 
are providing the starting point for new leaps of understanding 
of the subject. The key has proved to be deep surveys of the
same fields at every wavelength possible, so that as complete a
SED as possible can be determined for a large number of objects.
At radio wavelengths, the VLA at 20cm is the workhorse instrument.
The large field-of-view, high sensitivity and good spatial
resolution allow images with thousands of sources per pointing
and $\sim$1.5 arcsecond resolution. New computing algorithms
combined with the current generation of fast computers have unlocked
this potential of the VLA, which has been there for 25 years.

The $\mu$Jy radio sky gives unique information about the evolution of
faint, distant galaxies from the synchrotron emission observed from
both star-forming and AGN-dominated systems. This complements data
from other wavelength bands. The ratio of radio to FIR luminosity has
the potential to tell us whether star-formation or AGN are dominant 
\citep[e.g.,][]{Yun01}.
For star-forming systems the physical size of the emitting regions
and their relation to brightness distributions in other bands
can tell us about the extent of the star-forming regions independent of
dust extinction and perhaps can tell us if a wind is required. 
Radio AGN are dominantly  radio jets
and thus tell us about the mechanical energy flowing away from the
black hole, instead of the radiation dominated emission seen at
other wavelengths. Radiation and mechanical energy dominated AGN may
result from different phases of the black hole growth \citep{ch05,si07}
and thus it is important to study both types of AGN. The sensitivity
of this survey 
reaches a surface density of several sources per square arcminute and
thus allows us to
study a large sample of distant galaxies with only one pointing of
the VLA.
In this paper we report the results of the radio survey only. Other
related papers will discuss the observations at other wavelengths
and what we can learn by combining all the data on the objects in this
field.

\section{Selection of the Field}

The field for these observations was selected from the nearly 50
square degrees observed in the SWIRE legacy survey \citep{lon03}. The goal was
to select the best location in this survey for a deep radio field.
The Lockman hole area of the survey was chosen due to its uniquely
low HI column density and IR cirrus. This makes it ideal for the
Spitzer IR observations and followup in many other wavelength bands.
The declination of this region is also ideal for a VLA 20cm synthesis
since the current VLA 20cm system temperature increases below 50
degrees elevation and long tracks can be observed above this limit
in the Lockman hole. The NVSS survey \citep{con98} was searched for a
pointing center with a minimum number of strong sources which would
fall within the VLA primary beam and its first sidelobe. The region
was also required not to have sources as strong as a Jansky within 
a few degrees of the field center. The best field was centered at
the J2000.0 coordinates of 
10$^h$46$^m$00$^s$, 59\arcdeg01\arcmin00\arcsec.

\section{Observations, Reduction and Cataloging}

	Observations were made with the VLA in A, B, C and D
configurations for a total of almost 140 hours on-source between
December 2001 and January 2004.  In table~\ref{OR}, we summarize the 
parameters of the observing runs.
Since the total time is dominated by the A configuration, the final image
for analysis had a resolution $\sim$1.6 arcsecond.  The data
were all taken in spectra line mode 4, which provides seven
3.125 MHz channels in each of 2 IFs (centered at 1365 and
1435 MHz) and each of two polarizations. Five second
integration times were used in the A and B configurations and 10
seconds in the C and D configurations. The integration
times and channel bandwidths were chosen to minimize 
tangential and radial smearing of the images away from the
field center. This combination of parameters produces the best
compromise for imaging sensitivity and quality possible with
the current VLA correlator which dates from the 1970's. This
means that there is radial smearing of the image due to the
finite bandwidths of the channels which does degrade the
images away from the field center and which we will need to
take into account in the analysis of the image. Currently
the EVLA project is replacing the correlator, receivers and
associated electronics on the VLA. By the end of the project now
scheduled for 2012, much
wider bandwidths and narrower channels will allow more ideal
images as well as five to ten times higher sensitivity.
Another important consideration for deep
VLA work at 20cm with the present system is that the system
temperature increases as function of zenith distance due
to increased ground pickup due to the undersized VLA L-band feed.
 For this reason almost all the
observations were scheduled within 3.5 hours of source transit
to minimize this effect. This problem will also be dramatically
improved with the coming of the EVLA.

\subsection{Calibration \& Editing}

	All the radio reductions
use the AIPS package. Since the total
bandwidth is relatively large for a spectral line experiment,
for each daily observation we first split a bright point source 
calibrator from the raw database and applied a phase 
self-calibration. Then we used this otherwise uncalibrated
source to calculate a bandpass correction before we proceeded
with the rest of the regular continuum calibration. Regular
continuum calibration then continued, using the bandpass
calibration to flatten the spectral response, and thus
avoid the effects of the raw, sloping spectral response.
Calibration was made in the standard way to the Baars flux 
density scale \citep{baars} using 3C286 as the flux calibrator. The weights 
associated with each integration 
were also calibrated as a function of observed system temperature
as part of the standard AIPS processing.
	The calibrated dataset was first clipped using the
AIPS program CLIP well above the total found flux density found
in the field from lower resolution observations in order to remove
a few high points before beginning the self-calibration process.

\subsection{Imaging \& Self-calibration} 

The D-array data were first imaged to radii 8 degrees outside the
field center. The field was chosen to minimize the effect of
confusing sources; however, thirty-four confusing sources outside the 
primary beam were identified which needed to imaged outside the main
mosaic to optimize the final image. 
The most distant source which was included in the deconvolution is
1.8 degrees from the field center.

For the A-configuration data on the first day,
the entire primary beam was then imaged using the
AIPS task IMAGR and the 3D, multi-facet options. Besides
37 facets, each with $1000\times 1000$ points, to describe the primary
beam, another 34 facets, each with $500\times 500$ points, were centered
on all the confusing sources outside of the central region.
 	After initial imaging, each detected source had a tight clean box
placed around it to limit cleaning to real features. This procedure
allows each image to cleaned down to the 1$\sigma$ level and the
resulting clean components to be used in the self-calibration process. 
This process also eliminates almost all of the clean bias and does
not artificially reduce the noise on the image as happens when one
does not use boxes. 
	The images were self-calibrated, first just in phase,
and later in both amplitude and phase using the AIPS task
CALIB. These 71 images from the first day were then used as a 
fiducial image to ``self-calibrate'' the remaining 16 days of
A-array observations. The B, C, and D configuration observations
were also boot-strapped to the fiducial images in the same way.
Editing was carried before and during the self-calibration process
using both TVFLG and CLIP. Before self-calibration visibility points
with amplitudes much larger than the total flux density in the field
were removed with CLIP. After each self-calibration step the
clean model was subtracted from the calibrated data and high
residuals were deleted with TVFLG. Care was taken not to remove
moderately high points on the shortest baselines which might represent
data not well described by the model. 

     Since the data-volume from the 140 hours of observations is
so large and for the A and B configuration the same hour angles
were observed on several days, the next step was to compress the
data as much as possible before imaging. This was accomplished
using a multi-step process in AIPS which first converted all the
times into hour angles using TI2HA. Next the data for each
configuration were combined using DBCON and finally UBAVG was used
to average together visibilities with uv coordinates on each baseline close 
enough together to create negligible effect on the imaging. In practice
it is necessary to edit the date in the header for each day's data and
to perform the appropriate sorting to make this process work. However,
when the process is finished the full dataset was approximately ten
times smaller. This entire process, developed for this survey, has now
been combined into the AIPS procedure called STUFFR. 

The four configurations were then combined using DBCON in order to
make the final images. Of course, since the data were obtained over 
a period of more than a year, intermediate images were produced when
the data for each new configuration was obtained. Many experiments
were performed as to the best weighting, the optimum taper and the
imaging algorithm to use. In particular, multi-scale clean in IMAGR
was found to be most effective until the B configuration was added
to the A, C, and D data which were available earlier. However, for
the final image, a single resolution clean worked best. The box file
for the image was also improved many times during the process of
learning about the dataset. It was found that the best images were
produced with a set of boxes covering every source that could be
recognized in the field.

The images produced in this manor are still affected negatively 
by the pointing changes on outlying bright sources. This
is primarily due to the fact that the two circular polarizations
have different pointing centers ({\it i.e.} beam squint) on the VLA
due to the slightly
off-axis location of the feeds. This combined with the alt-az
geometry of the telescopes causes effective gain on a source far
off-axis to vary with hour angle. Also the slightly different
frequencies of the two IF's caused a slightly different primary
beam size and thus a source far from the field center has a higher
gain at the lower frequency.

As a final step to correct approximately  for these two effects, the
uv datasets were split into their separate IF's (2), polarizations (2)
and HA ranges (2). Then they were imaged and self-calibrated
separately. Finally the eight images of each facet were combined,
weighting each by $1/rms^2$ to form an optimum final image. The  37
facets covering the main lobe of the primary beam
were then combined into a single image using FLATN. As a final step
to produce the most uniform background MWF was used to flatten the
image further using the MODE over a $101\times101$-pixel support window.
This final image has a local {\it rms} of typically 2.7 $\mu$Jy over regions
100 pixels in diameter before correction for the primary beam. The
pixel size is 0.5 arcsecond and the clean
beam is $1.63\times 1.57$ arcsecond  at a position angle of 10 degrees.

\subsection{Cataloging}

	The inner $40\times 40$ arcminute region of the final image
was chosen for cataloging and further study. This region covers almost
the entire useful area of the primary beam. Since the bandwidth of
each channel smears the image radially by approximately $0.00223\times$
(the fractional bandwidth) $\times$ (the distance from the field
center), this effect needs to be taken into
account in analyzing the images. The smearing becomes comparable to
the clean beamsize 12 arcminutes from the field center. Furthermore,
as will be seen, the median size of the sources in the field is only
a little smaller than the clean beam. Thus for both instrumental and
physical reasons, in order to catalog the many sources as possible, it
is necessary to analyze the image with more than one resolution. This
was accomplished by convolving the full resolution to an effective
circular clean beam size of 3.0, 6.0 and 12.0 arcseconds. Each of the
four resulting images was then analyzed separately and the results
collated and compared. In practice, none of the sources had their best
detection in the 12.0 arcsecond image. 

	The AIPS program SAD was used for forming the initial
source lists down to peak flux densities. A catalog for each image
was formed down to a peak signal/noise (S/N) of 4.0. The residual images
from SAD were then searched to find any remaining sources  missed
by the program with a S/N greater than 5.0. For sources
with a S/N close to 5.0, the fitting process was repeated
by hand with JMFIT, using the local {\it rms} estimate over a
region 100 pixels in diameter. In this way a reliable list of sources with
S/N of 5.0 or greater at each resolution was compiled. Only sources
with peak S/N$\ge5.0$ in one of the three images (1.6\arcsec,
3\arcsec, 6\arcsec) are included in the final catalog. For both SAD
and JMFIT, the smearing due to the finite bandwidth was included in
the fitting process. 

The catalogs created this way were then collated. The relative
noise level varies for each source as a function of distance
from the field center,
which imposes
a selection effect on the final catalog which we discuss later. The
results for each resolution image were then
compared and the best fit was selected. If the peak S/N was more than 10\%
higher on the lower resolution image then that result was used. Finally,
if the total flux density of the source was significantly larger at lower
resolution then that result was adopted. In cases where the results
appeared inconsistent, the images were examined to resolve the issue.
If a source had a lower limit major axis from the best fit consistent
with zero, then the source was assumed to be a point and the peak
flux density is used as the estimate of the total flux density. For
sources only slightly resolved by this criteria, the total flux
density from the fit was cataloged. For unresolved sources, simulations have
shown that the fitted peak flux density for the unconstrained,
Gaussian functions fitted with JMFIT is the best estimate to the total
flux density \citep{g02}. For sources clearly much larger
than the clean beam, TVSTAT was used to estimate the total flux density
and the associated error given the number of beams included in the
sum. TVSTAT operates by the user drawing a region around the emission
region then summing the total signal in the region. Errors in the
flux density and position were calculated using the formalism of
\citet{con98}. A noise term proportional to the total flux density
of 3\% is included for each source to account for any calibration
errors and the uncertainty is in the primary beam correction. This
percentage error is consistent with the ratios of our flux densities
to the NVSS flux densities of sources $> 5$ mJy which are in the field.

\section{Results}

\subsection{Radio Catalog}

\begin{figure}[p]
   \epsscale{0.92}\plottwo{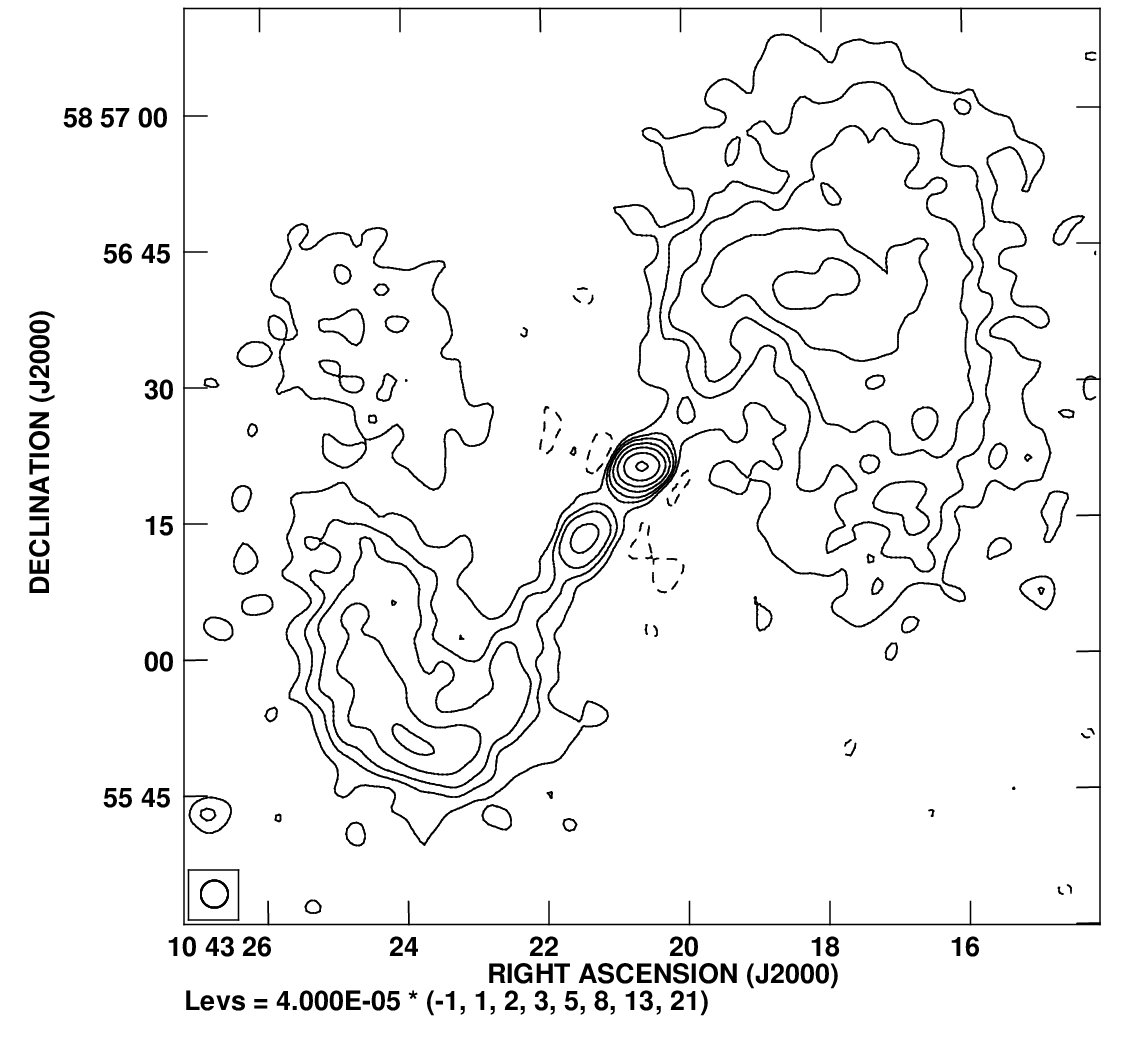}{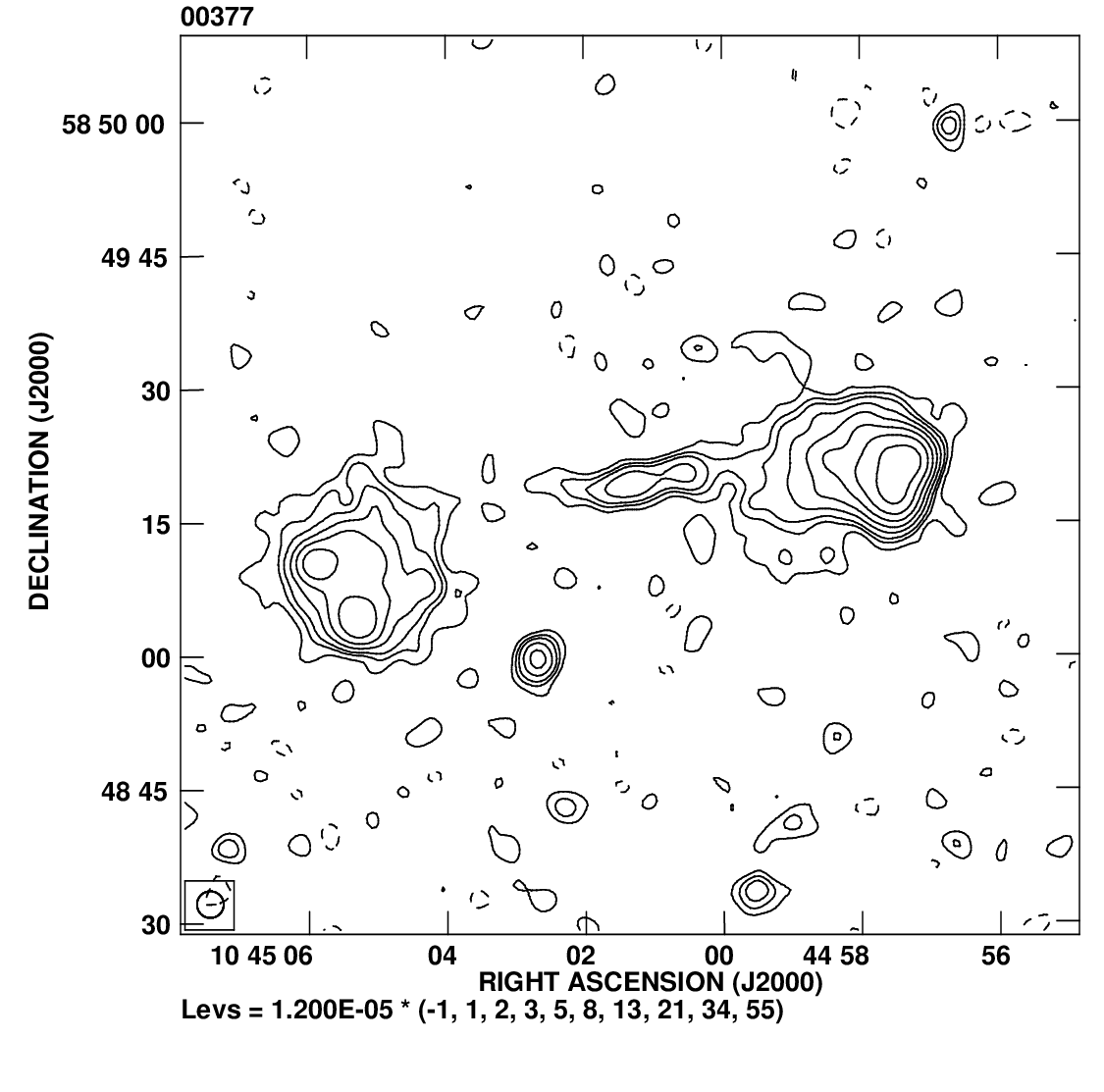}
   \epsscale{0.92}\plottwo{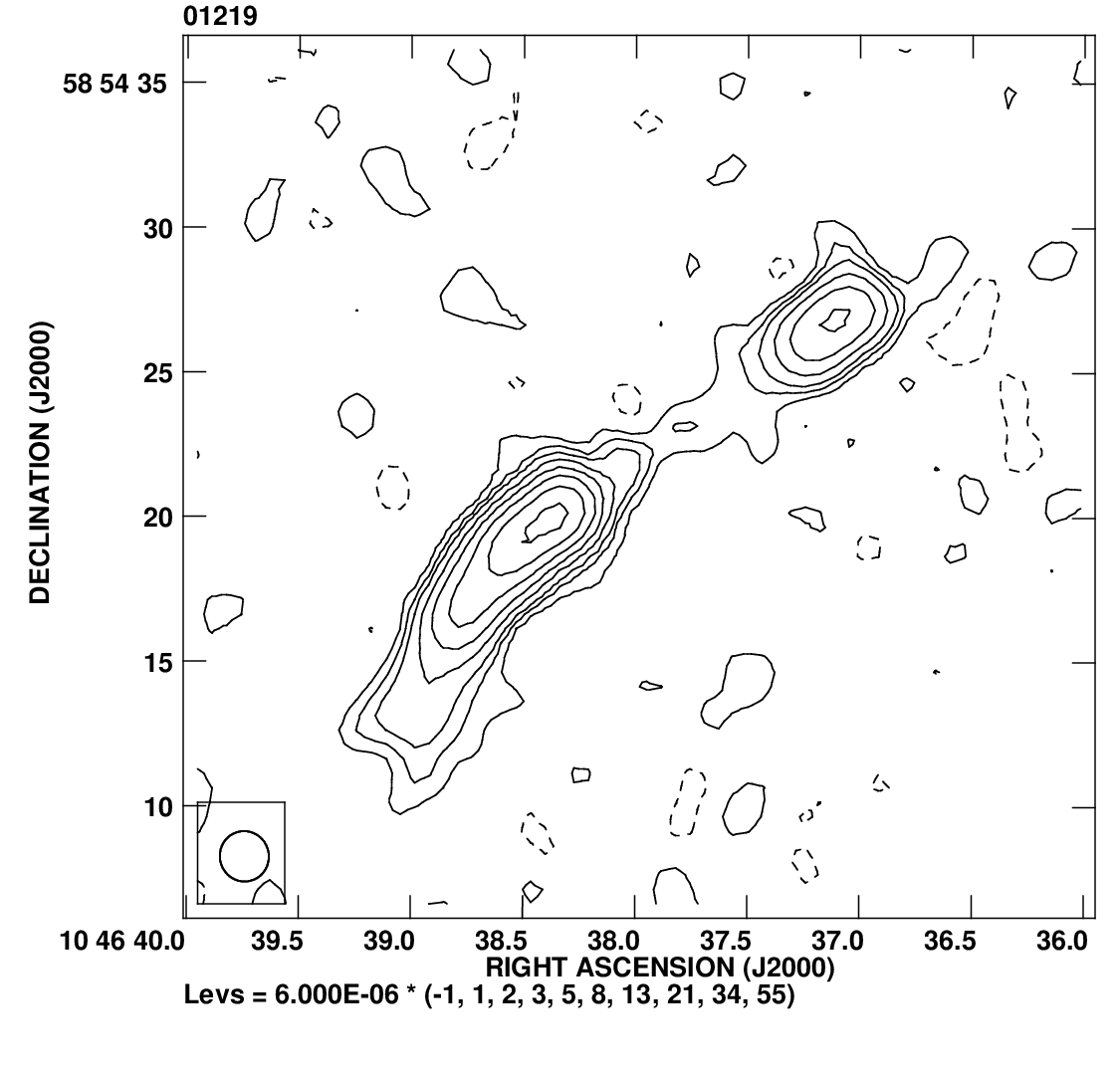}{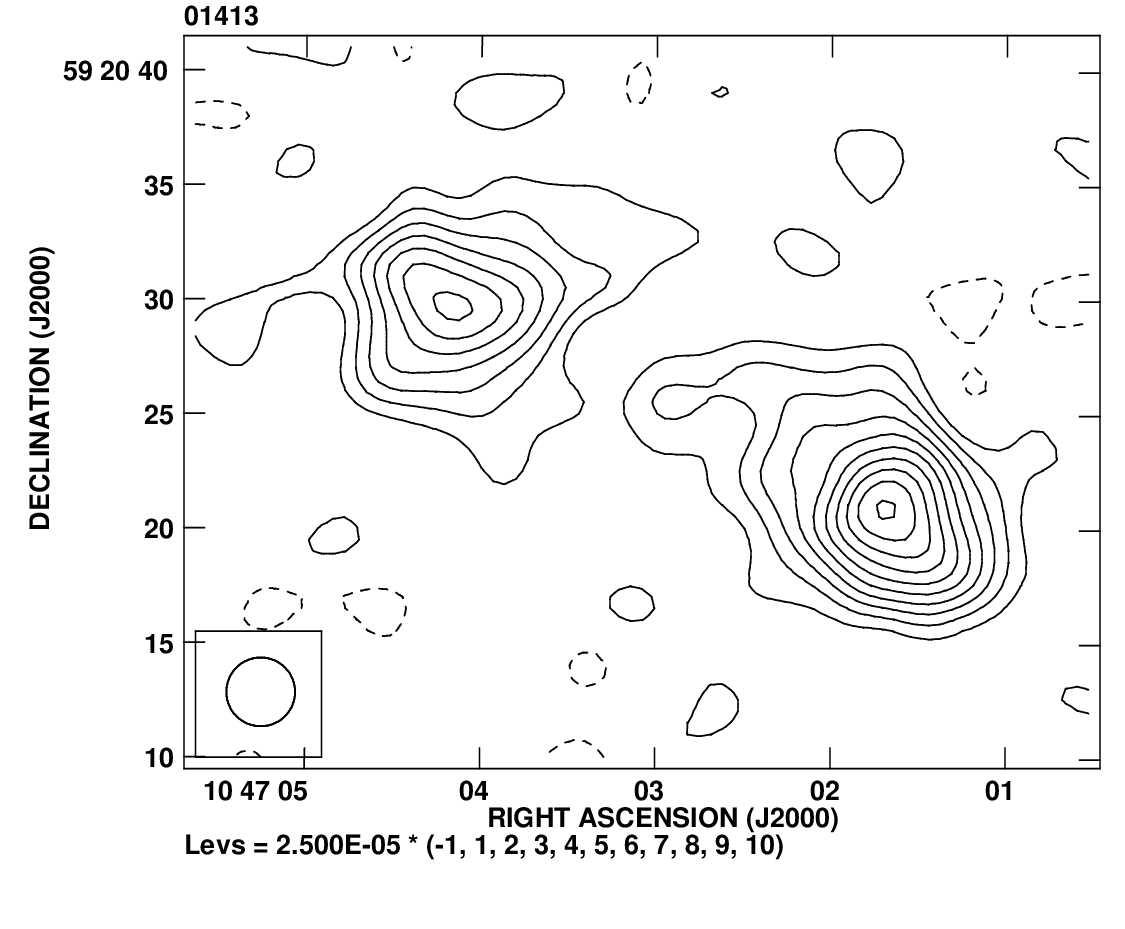}
   \epsscale{0.92}\plottwo{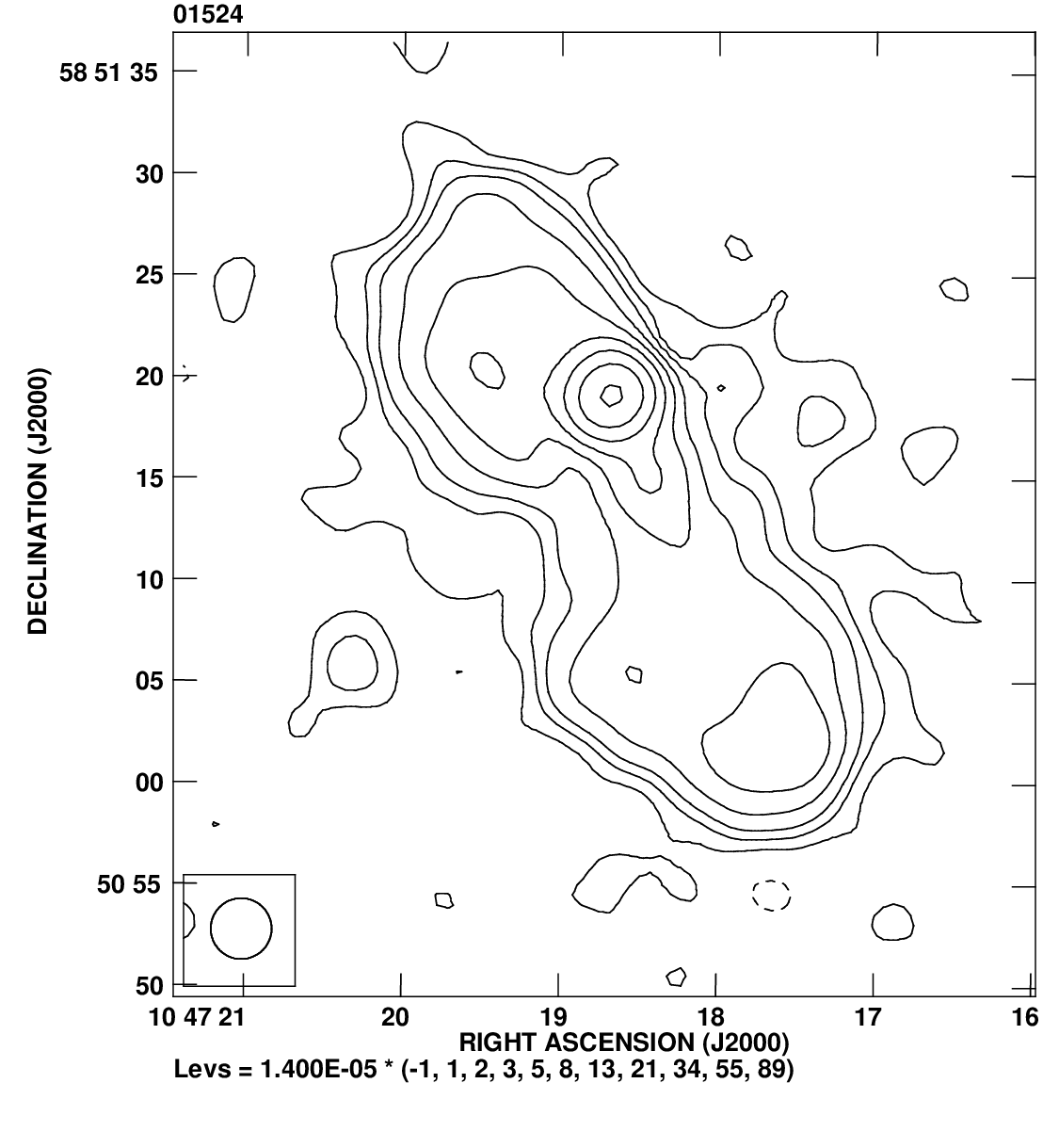}{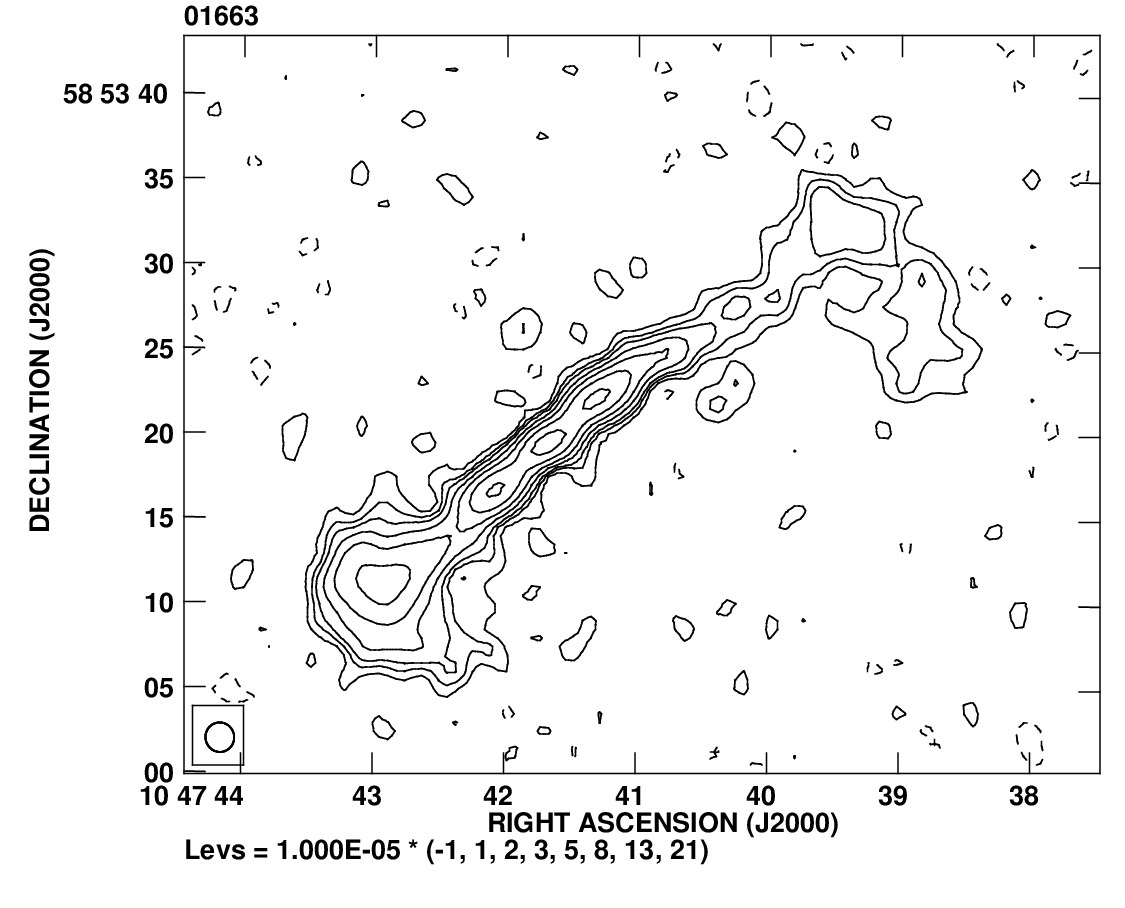}
   \caption{Extended Sources--Contour Levs in Jy \label{kntr2}}
\end{figure}

	In table~\ref{SL}, we give the first ten lines of the
radio catalog, while the rest is provided electronically.  Column (1) 
contains the source number. If the source has
a number less than 2000, then it was found in the original list found
with a S/N $\ge 5.0$ from running SAD on the full resolution image.
Sources with numbers $\ge 10000$ were found in lower resolution images
or in checks
of the residual images.
Columns (2) and (3) contain the radio RA and Dec along with the
estimated error. Column (4) contains the
observed (uncorrected) peak flux density from the map in $\mu$Jy per
beam. In column (5) we list the corrected total flux density and
estimated error. Column (6) contains the peak S/N. The error for
column (4) can be recovered by dividing column (4) by column (6). We
give the S/N as opposed to the error since the S/N was used to
define the catalog cutoff and later is used in the calculation of
log N- log S. In column (7), we give the best fit deconvolved size in
arcseconds. If a resolved two dimensional
Gaussian was the best fit, we give the major and minor axis size
(FWHM) and the position angle. Upper limits are given for sources
which were unresolved based on the results of JMFIT or SAD.  For
sources with very large sizes, as shown in
figure~\ref{kntr2} and \ref{kntr1}, sizes and total flux densities
were estimated directly
from the images using the AIPS tasks, IMVAL and TVSTAT.

\subsection{Angular size distribution}

This deep survey contains information about the angular size distribution 
as a function of flux density. However, our limited survey brightness
sensitivity means that we will miss larger sources near the survey
limit. We also have limited ability to determine source sizes for
sources much less than one arcsecond in size. However, our
Gaussian fits show that we resolve about half the sources we detect.
This allows us to give a good estimate of the median source size.
In a later paper we will discuss the
linear size distribution. However, just from the radio image the
angular size distribution gives us a hint about the nature of the
faint source population. Also as we survey to deeper limits we can
learn more about the interesting angular sizes for future radio
instruments as well as whether we may reach a natural confusion
limit where the sources themselves may overlap and thus limit our
ability to study fainter sources. 

For the purposes of this investigation we will limit ourselves to the
inner part of the image where the image is most sensitive. For the
fainter sources, we stay close
to the field center and for the relatively rarer, brighter objects we will
use a larger area. Sources
within five arcminutes of the field center have very small smearing due
to the bandwidth effects discussed earlier and the sensitivity
over this region is almost unaffected by the primary beam. Of course,
there is still a bias against detecting larger sources close to the 
S/N limit. For brighter sources, the higher S/N makes up for the
somewhat larger smearing, so we can measure their sizes more
accurately. However, their number density decreases, so we need a
larger area for adequate statistics. Table~\ref{stats} gives statistical 
properties of each sample subset we analyzed. In column 1 we give
the radius from the pointing center in arcmin; columns 2 and 3 contain
the minimum and maximum flux densities for the subsample; in column
4 we list the number of sources; columns 5 and 6 contain the mean
size and its error from the Kaplen-Meier analysis and column
7 contains the observed median size (uncorrected for incompleteness);
in column 8 we list the percentage of resolved sources which went into
the Kaplen-Meier analysis. 

In figure~\ref{ahist} we show histograms of the Kaplan-Meier estimate
of the distribution of fitted Gaussian major axis size versus flux 
density, using the IRAF task KMESTIMATE for this purpose. The IRAF
task allowed us to include the upper limits in the calculation,
although most of the sources that we used for this purpose have
non-zero detections for the major axis. The primary purpose of
this analysis is to determine how the median source size
evolves with flux density. Except for the lowest flux density bin for
which we quote an upper limit, significantly more than half the sources 
in each bin are resolved, but
in general there are some sources with upper size limits which exceed
the median. The Kaplen-Meier formalism provides a way to do a small 
extrapolation to the median taking into account the upper limits in a
consistent way. 
We then can compare this result with source size
distributions found from higher resolution observations. We show
the full results of the Kaplen-Meier estimates which are consistent
with the data trends but may underestimate the number of very small 
sources. 
The histograms show that the maximum source size we detect increases
with flux density. Below 100 $\mu$Jy the
cutoff is roughly the largest size we might expect to be able to
detect based on surface brightness, although this result depends
in detail on the source shape. Thus below 100 $\mu$Jy we suspect that
we are incomplete for the larger sources, especially below 30 $\mu$Jy.
Even if we ignore this effect, the median size for the sources we
detect remains above one arcsecond and is larger if we try to correct
for the missing population of larger sources. In the 20-30 $\mu$Jy
bin the median size even appears to increase to 1.5 arcsecond without
any corrections. These results are consistent with \citet{m05} and
suggest that the size distribution of the radio population may not
continue to decrease below 100 $\mu$Jy but may level out at a median
size $> 1$ arcsec.

\subsection{Log N - Log S}

It might seem easy to calculate the density of sources as a function
of flux density from a catalog such has been compiled in this paper.
However, several effects contribute to make the source list incomplete
even at fairly large S/N. The detection rate is obviously directly
affected by the primary beam attenuation. However, both the finite
bandwidth of each channel (3.125 MHz) and to a lesser extent,
the finite time averaging smear the sources as a function of radius
from the field center and thus reduce the peak amplitude (see figure
2 of \citet{f06}). Furthermore, the fact that the sources have a
median source size on the order of the clean beam size, also reduces
the completeness of the survey. Thus besides counting the sources as
a function of flux density, an incompleteness correction needs to be
included in the calculation. Since we do not know the actual
distribution in angular sizes as a function of flux density, this
correction can only be approximate and based on the properties of
sources at higher flux densities. However since we find evidence in
the previous section that the median sizes are consistent with being 
approximately constant as a function of flux density, we will make the
simple assumption that the distribution of angular sizes remains
constant at flux densities below 1 mJy. 

One problem is that most very deep surveys have been carried out using
only the VLA's highest resolution at 20cm to avoid confusion and obtain
the best possible positions. However, such surveys select against
finding large sources and thus our view of the radio sky is
biased against large sources. Our survey, while still having
some such bias, used all four VLA configurations and thus has
the necessary spatial frequencies to detect quite large sources.
As outlined earlier we did search for large sources using restoring
beams as large as twelve arcseconds. We also are more sensitive than
any previous survey, so we are more sensitive to large sources
than other VLA deep surveys. However, due to confusion and our
overall weighting toward longer baselines, we are not as sensitive
to large sources of a given flux density as we are to small
sources.  

\citet{f06} finds that $8\pm 4$\% of $\mu$Jy sources are larger than 
4 arcsec. In our $300-1000$ $\mu$Jy sample we find that $5/30$ or
about 17\% of our sources are bigger than 4 arcsec. Even in our
$30-100$ $\mu$Jy sample, we find $4/50$ sources are bigger than 4
arcseconds and none are detected which are bigger than 6 arcsec. Assuming 
that the size distributions are the same as a function of flux
density, we should have missed some larger sources in this bin.
Thus it seems likely that at least 15\% of the
sources in our field are bigger than 4 arcseconds. Overall, we find
slightly larger sources than \citet{f06} and we suspect we might be
still underestimating the source size distribution.

\subsubsection{Sensitivity vs Distance from the Field Center}

 In their figure 2 \citet{f06} show the sensitivity decreases as a
 function of distance
from the field center of a point source at our full resolution due to the 
VLA primary beam, the bandwidth smearing due to our 3.125 MHz channels and
the time smearing. The latter two effects are due to the smearing of
the synthesized beam by our finite time and frequency resolution. 
These effects can be mitigated by using a bigger synthesized beam. This
mitigation is achieved in practice by convolving the cleaned image by the
appropriate two dimensional Gaussian to produce the desired new beam size.
However, this process increases the noise on the image, so picking the
best beam size is a trade-off which changes with distance from the field
center. In figure~\ref{rel20}, we show the relative point-source sensitivity 
for the three different resolutions we analyzed. Note that the three arcsecond
(red) curve crosses the full resolution (blue) curve about nine arcminutes
from the field center and thus beyond this radius the three arcsecond image is
actually more sensitive than the full resolution image for point sources.

A further complication is that the sensitivity for resolved sources
of different sizes is also a function of beamsize and distance from the
field center. In figure~\ref{size137} we show examples for three different 
Gaussian source sizes as an example (one, three and seven arcsecond FWHM,
thin solid lines, dashed lines and thick solid lines respectively). 
The one arcsecond sources, the full resolution and three arcsecond resolution
curves now cross about 6.5 arcminutes from the field center. For three
arcsecond sources, the red (three arcsecond resolution) curve is always
above the blue curve. For as large as seven arcsecond, the green six arcsecond 
curve is always best.

 Since most of the sources are large 
enough to have images significantly bigger than a point source at our
highest resolution even in the field center,
the clean beam which is most sensitive varies with both distance from
the field center and source size. Furthermore depending on the source
size we are incomplete at each flux density and peak S/N due to
the source size distribution. Since we do not know this distribution in
great detail we can only correct for this affect approximately. In practice,
there seem not to be many sources larger than, say, five arcseconds, thus
above some flux density we don't miss enough sources to affect the source
counts significantly for most purposes. However, at the bottom of the
catalog, where the information is most interesting, we must be somewhat
uncertain due to not knowing the source size distribution in detail.

\subsubsection{Calculating and Correcting Log N - Log S}

We first calculate the number counts ignoring the source size distribution.
For
each detected source we calculate the area on the sky over which it
can be detected, taking into account the three different resolutions we
have used for our cataloging process.
In figure~\ref{sa13comp} we show the log N - Log S results for our
survey (blue points) and the results of the shallower SA13 survey \citep{f06}
using yellow squares.
One can see that both curves turn down at low flux densities but not
at the same flux density. This certainly suggests some sort of incompleteness.
In order to explore this result, we calculated our log N - Log S 
for different S/N cutoffs ranging from 5 to 10. These results are
shown in figure~\ref{rawlogNS}. One can see that the higher the S/N
the earlier the higher the flux density at which the curves turn down. 
 The SA13 curve is in good agreement with our S/N cutoffs which
 approximately match
the \citet{f06} flux density cutoff. This seems to suggest that the
incompleteness in the full catalog comes from some other source than S/N. 

At least part of this effect is due to the fact that we are
resolving the sources, as discussed in the last section. In order to 
correct for this effect we need to know the source size distribution
in some detail. Clearly we have only a rough idea of the size
distribution as a function of flux density. In order to correct
for source resolution, we have assumed a size distribution based
on the higher resolution results of \citet{m05} at small source sizes
and a large 
source-size tail based on our results (figure~\ref{angsize}), which should be
more complete because of our better {\it uv} coverage on short
baselines.
We assume this distribution is the same for all flux densities
we observe. 

 Using this assumed size distribution,  
we can then correct for our incompleteness as
a function position and resolution of our images. In figure~\ref{corlogNS}
we show the effect of this correction. The impact is to flatten the
counts. The scatter in the estimated source densities is primarily
due to the interaction of the size correction and the inverse areas
with the S/N cutoffs. If a few sources lie close to the S/N cutoff
then they can dominate the estimate and the associated error. In order
to minimize this effect in the estimate of log N - log S, we have
calculated the corrected log N - log S with cutoffs from S/N=5 to
S/N=10 in steps of 0.2 in S/N. Then we have performed a weighted
sum of the estimates in each flux density bin. The results are
summarized in table~\ref{counts} and  shown
in fig~\ref{final}. In this plot we can see a flat distribution with
flux density for 
normalization coefficients of the differential log N - log S. 
 In order to estimate the errors in the new process described above we
constructed 100 Monte-Carlo simulations with similar log N - log S and
the parent angular size distribution we assumed for the analysis.  Using our
algorithm described above we then used the
simulations to estimate the one sigma error bars shown in 
figure~\ref{final}. 
Our result is still subject to errors in our assumed parent
size distribution and any cosmic variance; we have not attempted to
include these potential sources of error in our error estimate. 

The resulting distribution, although dependent on the assumed angular size
distribution, is consistent with a flat shape for the normalizing
coefficient of the differential log N - log S distribution. Thus we
find no evidence that the counts are turning down when one takes into
account the size distribution and other effects affecting the
observations. Since our observed  size distribution down to the faintest
levels in the survey  is consistent with our assumed global size distribution,
the results in figure~\ref{final} and table~\ref{counts} are our best
estimate of log N - log S. We have
searched our images for sources both with an automatic program and
then verified the results by searching the residual images by
eye and refitting each source down to a peak S/N of sigma with JMFIT.  Thus we
do not think we have missed many, if any, five sigma sources. If
we have missed sources then our results are underestimates of
the counts at the faint end. We see no evidence for a turn-down
in the counts or a decrease in source size.

\subsubsection{Comparison with previous log N - log S determinations}

In figure~\ref{lognsall} we show our new values compared with other
recent determinations of log N - log S for $S_{20cm} < 1$ mJy. 
Our approach to cataloging and estimating log N - log S is different
in detail compared to previous studies.
 We believe it takes into account the best estimate of the
size distribution in a better way than has been used for the other
data shown in figure~\ref{lognsall}. Furthermore, our result is for
a single direction in the sky and is, of course, subject to cosmic
variance. Nonetheless, while our points are a little higher than the
other results, the function is generally flat from 1 mJy down to our
lowest point at 17$\mu$Jy. At our survey limit, counting the inferred
resolved population, there are $\sim6$ sources per square arcminute. 

Since there is no trend toward smaller sources at lower flux density,
this raises the issue of whether we are approaching the natural
confusion limit. If the counts remain flat and the size distribution
remains the same, we are less than an order of magnitude from this
limit. With the EVLA it should be be possible to improve the limit
by this factor and test whether we are approaching this ultimate limit
and whether even deeper surveys with a hypothetical SKA will suffer
for this fundamental problem. However, the details of the radio source
structure remain interesting even if the natural sources confusion is
approached. Since the typical source is $\sim$1 arcsecond in size,
this argues for any next generation continuum survey array
to be designed for high spatial resolution, perhaps $\sim0.1-0.01$
arcsecond and
to concentrate on source structure and its spatial relation to high resolution
results at other wavelengths, not deep detection surveys.

\section{Conclusion}

We have presented the deepest image so far of the radio sky in terms
of source density and provided an electronic catalog for sources
with peak signal-to-noise ratios $> 5$.  The median size distribution
for these sources continues to be $\sim1.2$ arcseconds or a bit larger 
down to the bottom of the catalog.
After a correction for incompleteness due to source size, we find that
the log N - log S normalization factor remains approximately flat down
to 15$\mu$Jy, corresponding to about six sources per square
arcminute. If this trend continues the 20cm natural confusion limit
may be reached near one $\mu$Jy.

\clearpage
\begin{deluxetable}{lrrrrrlrc}
\tablecolumns{9}
\tablewidth{0pt}
\tablecaption{Observing Runs Summary\label{OR}}
\tablenum{1}
\pagestyle{empty}
\tablehead{
\colhead{Configuration} & 
\colhead{Startdate} &
\colhead{Enddate} &
\colhead{Hours}}
\startdata	
A&02Jan27&02Mar11&104.0\\
B&03Dec15&04Jan14&27.5\\
C&03Jan06&03Jan06&6.5\\
D&01Dec15&01Dec15&1.6\\
\enddata
\end{deluxetable}
\clearpage
\begin{deluxetable}{lrrrrrl}
\tablecolumns{7}
\tablewidth{0pt}
\tablecaption{Radio Source Catalog\label{SL}}
\tablenum{2}
\pagestyle{empty}
\tablehead{
\colhead{Name} & 
\colhead{RA(2000.0)} &
\colhead{Dec(2000.0)} &
\colhead{Peak} &
\colhead{Total} &
\colhead{S/N} &
\colhead{Size}\\
\colhead{} &
\colhead{} &
\colhead{} &
\colhead{$\mu$Jy/b} &
\colhead{$\mu$Jy} &
\colhead{} &
\colhead{arcsec}}
\startdata
00004&10 43 19.29(0.04)&59 08 42.6(0.3)&   35.7&   394.4(43.1)&13.7& 2.1x0.4p=109.\\
00005&10 43 20.06(0.07)&58 51 42.3(0.6)&   14.1&   141.9(27.0)&5.4&$<$3.5\\
00006&10 43 20.63(0.57)&58 56 20.8(4.4)&  227.3&  42293.0(4343.5)&71.0&146.\\
00007&10 43 20.84(0.05)&58 53 50.4(0.4)&   27.3&167.7(21.5)&21.5&$<$3.3\\
10007&10 43 21.84(0.03)&59 12 54.4(0.2)&   24.4&   213.7(30.1)&7.6&$<$3.1\\
00010&10 43 23.44(0.02)&59 17 50.8(0.2)&   29.9&   482.7(58.0)&9.2&$<$2.7\\
00012&10 43 25.05(0.02)&59 06 16.1(0.2)&  108.4&   653.5(43.6)&33.8&1.7x1.3p=45.\\
00013&10 43 26.79(0.06)&59 01 47.7(0.5)&   32.0&   330.2(37.4)&12.4&2.6x1.4p=81.\\
20013&10 43 26.81(0.02)&58 55 43.0(0.2)&   25.1&   123.0(16.9)&7.8&$<$3.4\\
00015&10 43 28.02(0.02)&59 13 05.8(0.2)&   27.2&   211.4(27.4)&8.4&$<$2.4\\
\enddata
\end{deluxetable}
\clearpage

\begin{deluxetable}{rrrrrrrcl}
\tablecolumns{8}
\tablewidth{0pt}
\tablecaption{Source Size Summary\label{stats}}
\tablenum{3}
\pagestyle{empty}
\tablehead{
\colhead{Radius}& 
\colhead{Minimum}&
\colhead{Maximum}&
\colhead{Sources}&
\colhead{Mean}&
\colhead{Error}&
\colhead{Median}&
\colhead{Resolved}\\
\colhead{arcmin}&
\colhead{$\mu$Jy}&
\colhead{$\mu$Jy}&
\colhead{}&
\colhead{arcsec}&
\colhead{arcsec}&
\colhead{arcsec}&
\colhead{\%}}
\startdata	
5&13&20&21&0.9&0.2&$<$1.5&19\\
5&20&30&36&1.6&0.1&1.5&58\\
5&30&100&51&1.6&0.2&1.1&75\\
10&100&300&29&2.3&0.6&1.0&72\\
20&300&1000&30&2.6&0.7&1.1&83\\
\enddata
\end{deluxetable}
\clearpage

\begin{deluxetable}{rrr}
\tablecolumns{3}
\tablewidth{0pt}
\tablecaption{Differential normalized source counts for 1046+59. The
  table contains 1) $S_l$ (the lower flux density limit of the bin), 
2) $S_h$(the upper flux density limit if the bin), 3) the normalization
  factor and its error. \label{counts}}
\tablenum{4}
\pagestyle{empty}
\tablehead{ 
\colhead{$S_l$}&
\colhead{$S_h$}&
\colhead{$S^{2.5}dN/DS$}\\
\colhead{$\mu$Jy}&
\colhead{$\mu$Jy}&
\colhead{Jy$^{1.5}$sr$^{-1}$}}
\startdata	
15&19&$6.48\pm 1.51$\\
19&24&$6.50\pm 0.89$\\
24&36&$6.11\pm 0.58$\\
36&54&$5.75\pm 0.69$\\
54&100&$7.01\pm 0.74$\\
100&150&$6.24\pm 1.11$\\
150&300&$4.46\pm 0.94$\\
300&1500&$4.23\pm 1.26$\\
\enddata
\end{deluxetable}
\clearpage

\clearpage

\begin{figure}[p]
 \epsscale{0.74}\plottwo{20CM_00671.ps}{20CM_00727.ps}
 \epsscale{0.74}\plottwo{20CM_01081.ps}{20CM_01738.ps}
  \caption{Extended Sources--Contour Levs in Jy \label{kntr1}}
\end{figure}
\clearpage

\begin{figure}[p]
 \epsscale{0.9}\plottwo{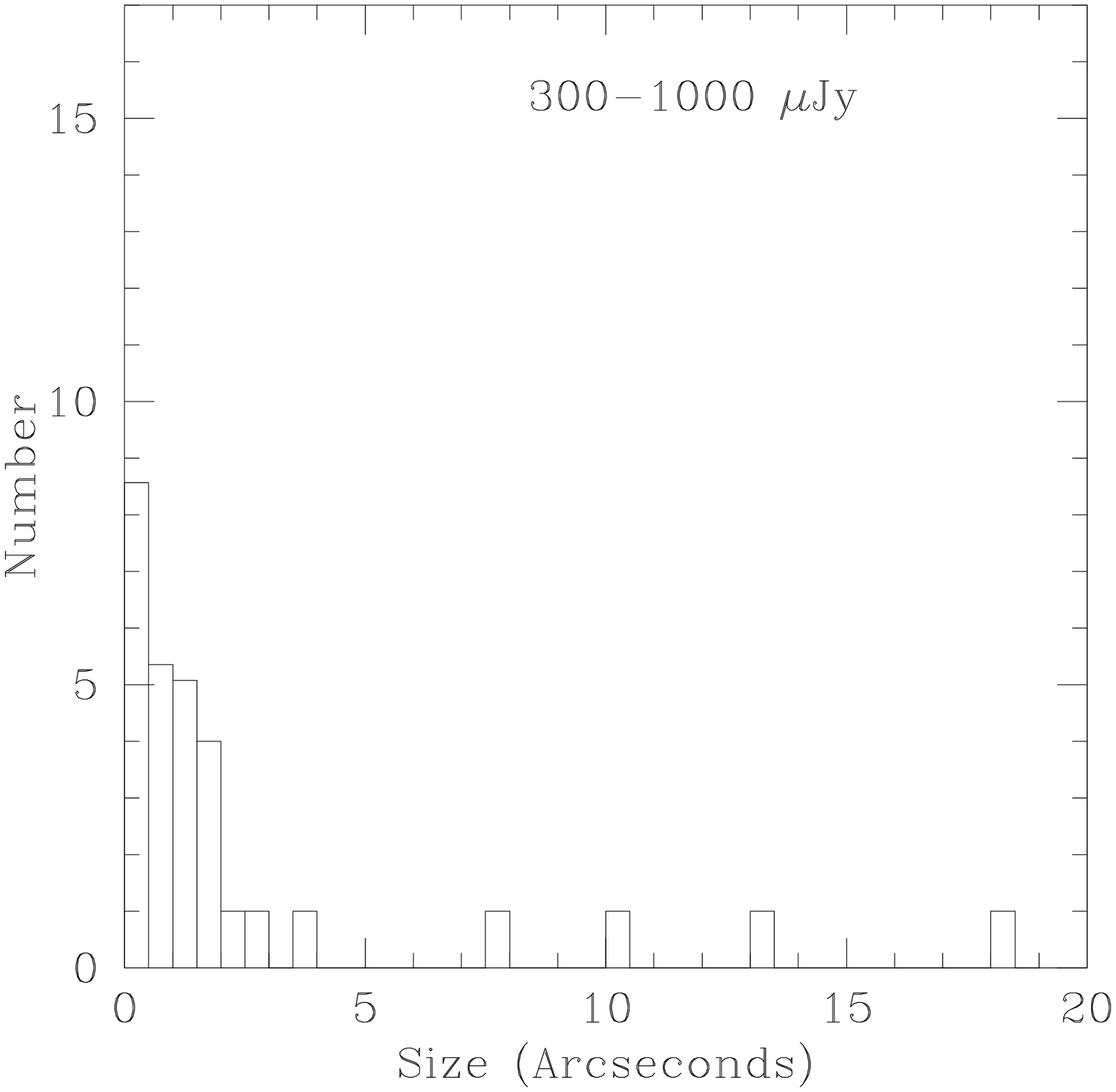}{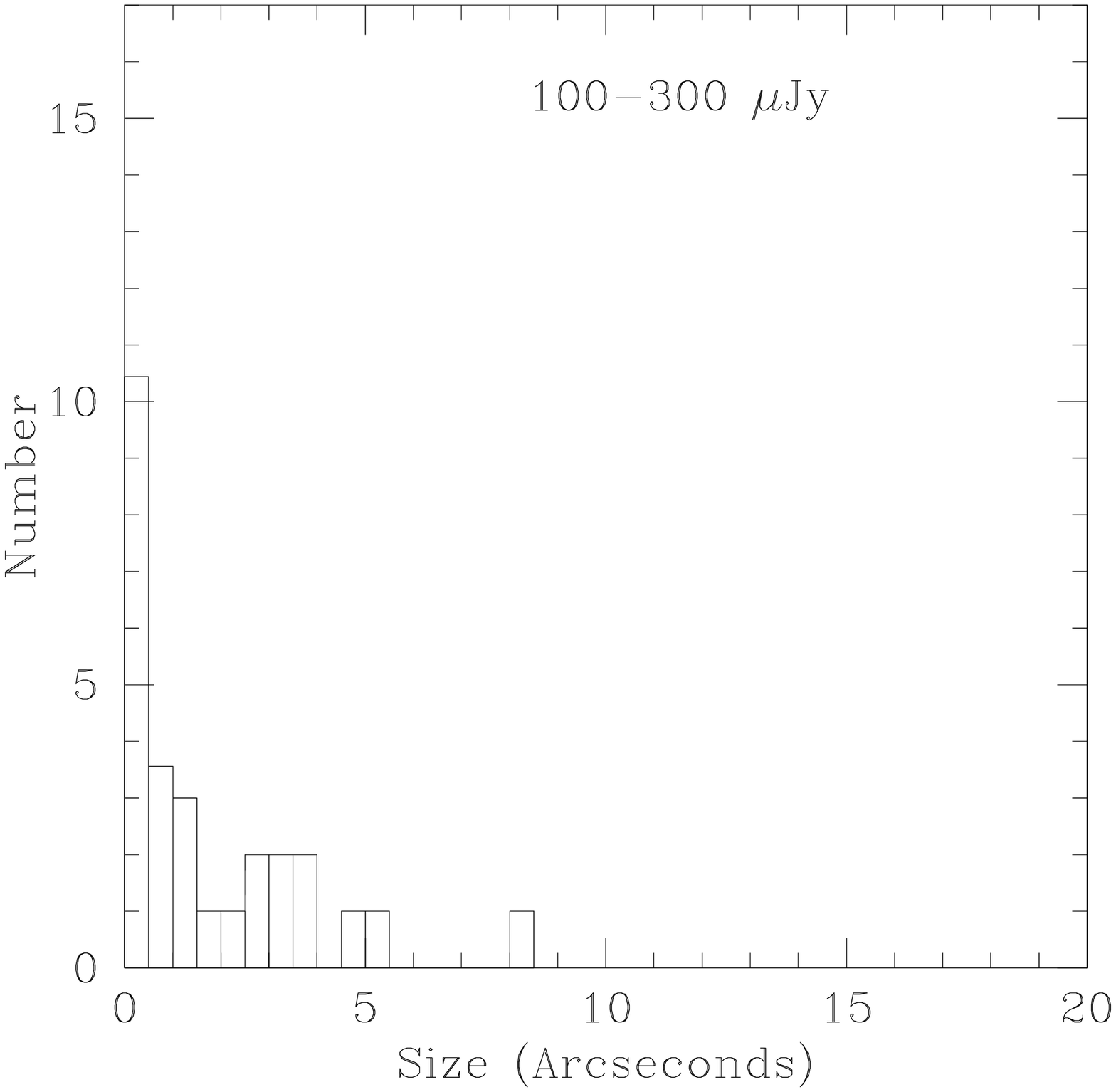}
 \epsscale{0.9}\plottwo{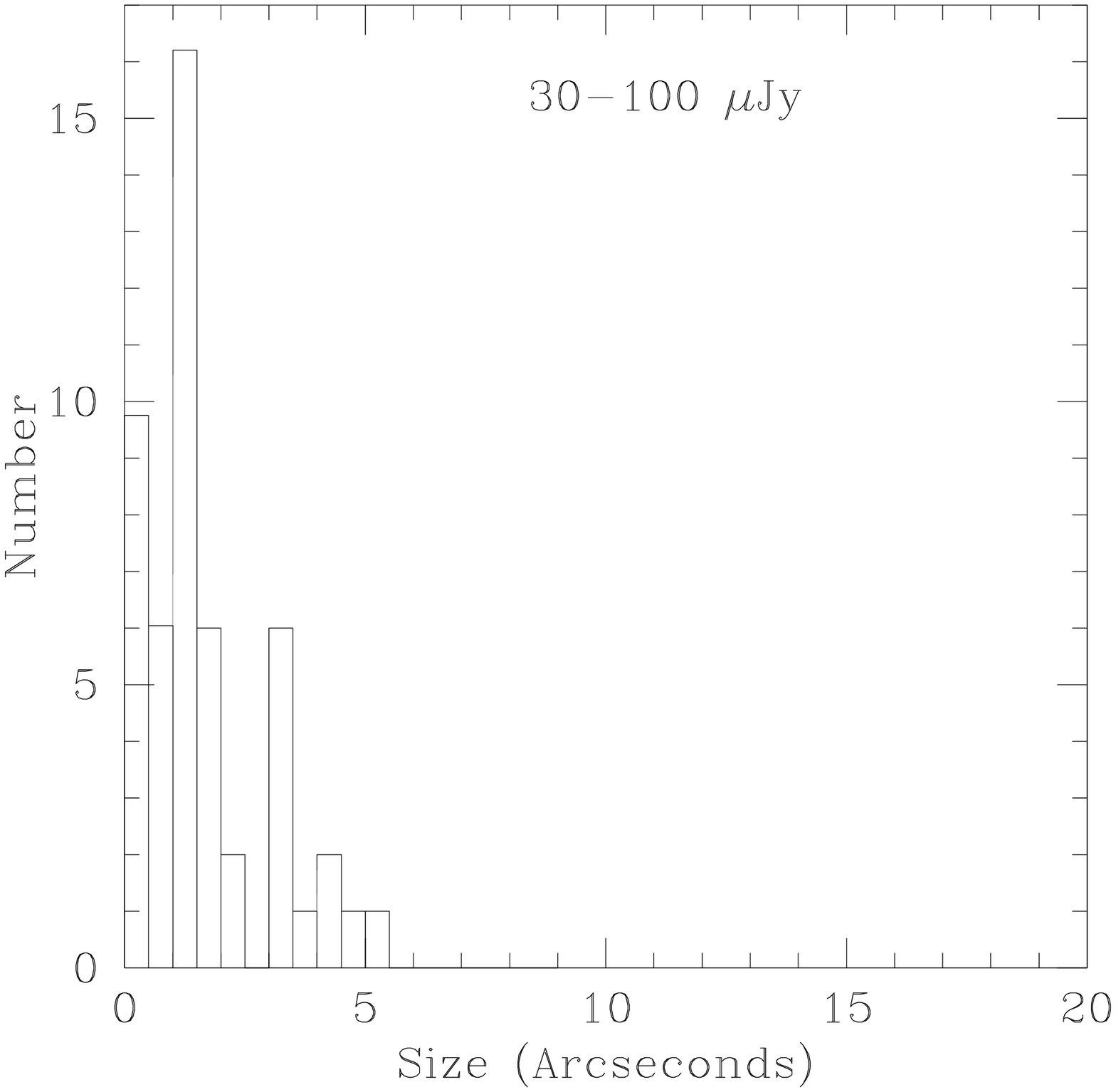}{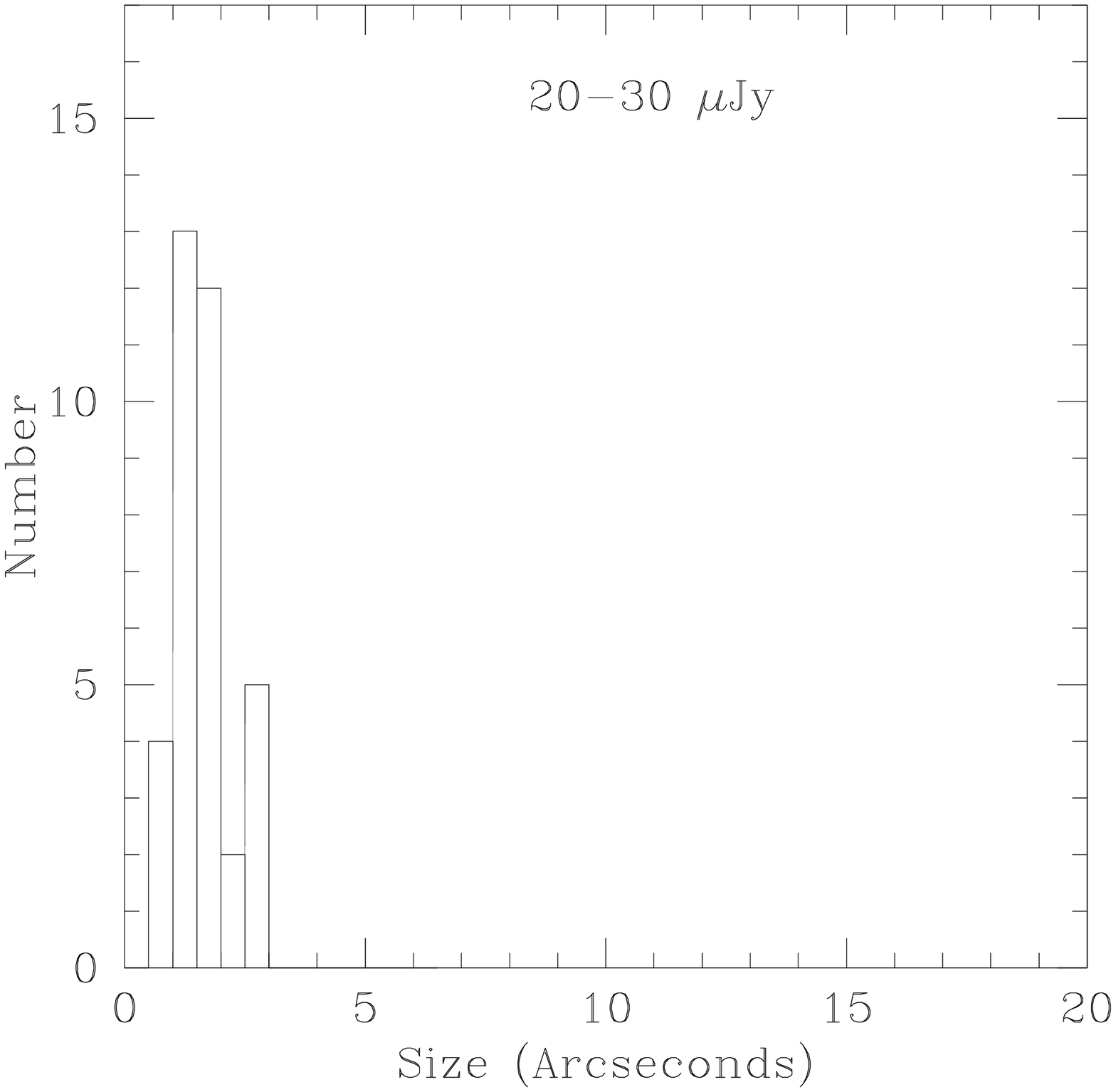}
 \epsscale{0.9}\plottwo{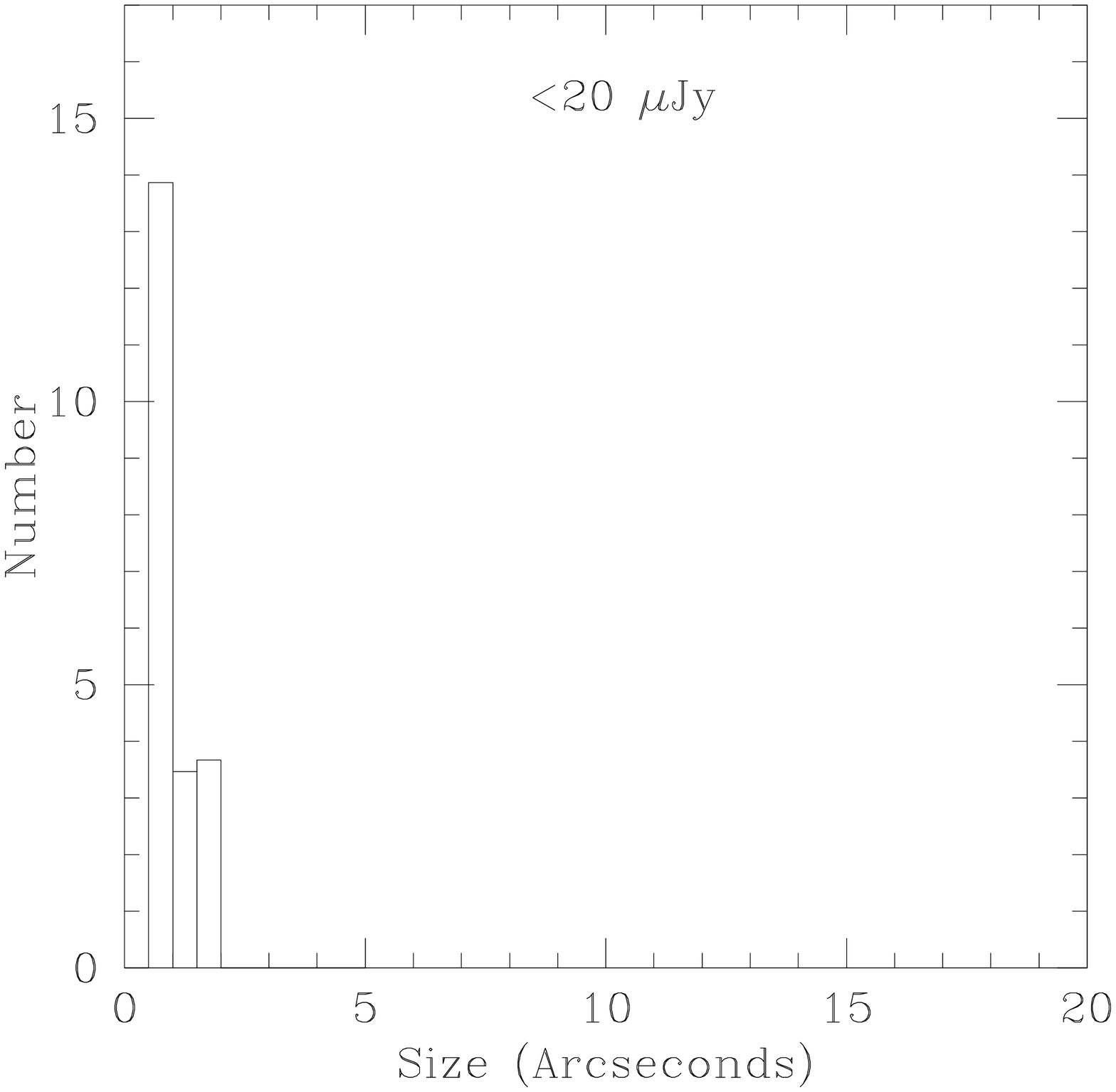}{blank.ps}
 \caption{Histograms of the Kaplan-Meier estimates of the deconvolved 
angular-size vs flux density for various ranges in flux density. Note
that the Kaplan-Meier formalism may have underestimated the number of
very small sources where it is poorly constrained but likely gives a
good estimate of the median size.  \label{ahist}}
\end{figure}
\clearpage


\begin{figure}
\plotone{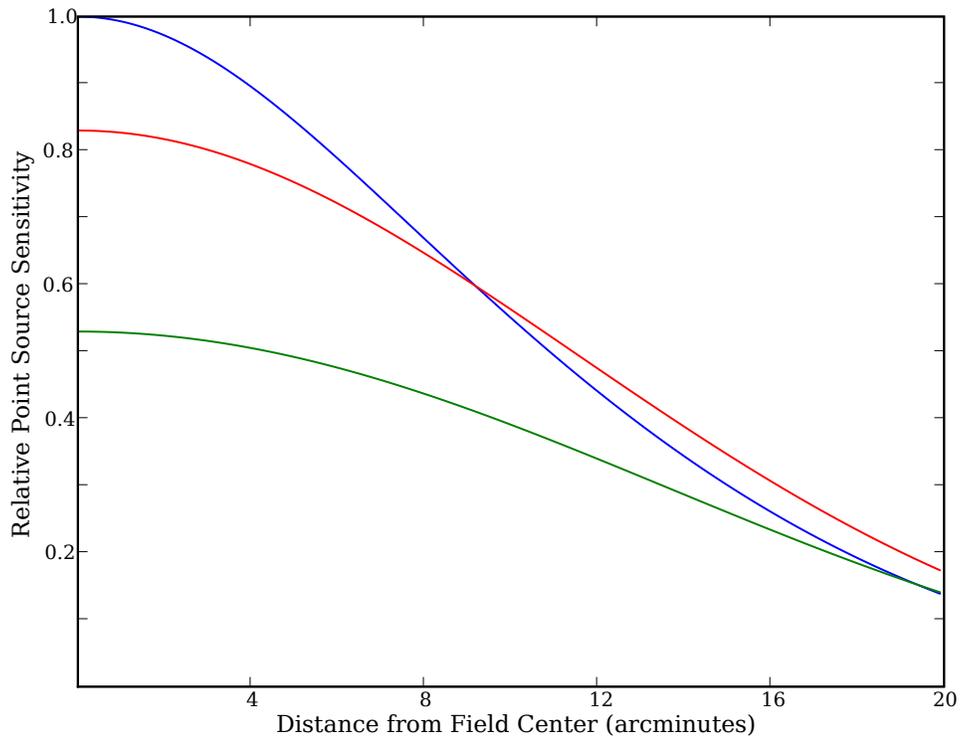}
\caption{Relative point-source sensitivity as a function of distance
from the field center for the three image
resolutions used for cataloging: blue: full resolution ($\sim 1.6$\arcsec),
red ($3.0$\arcsec ), green ($6.0$\arcsec). \label{rel20}} 
\end{figure}
\clearpage

\begin{figure}
\plotone{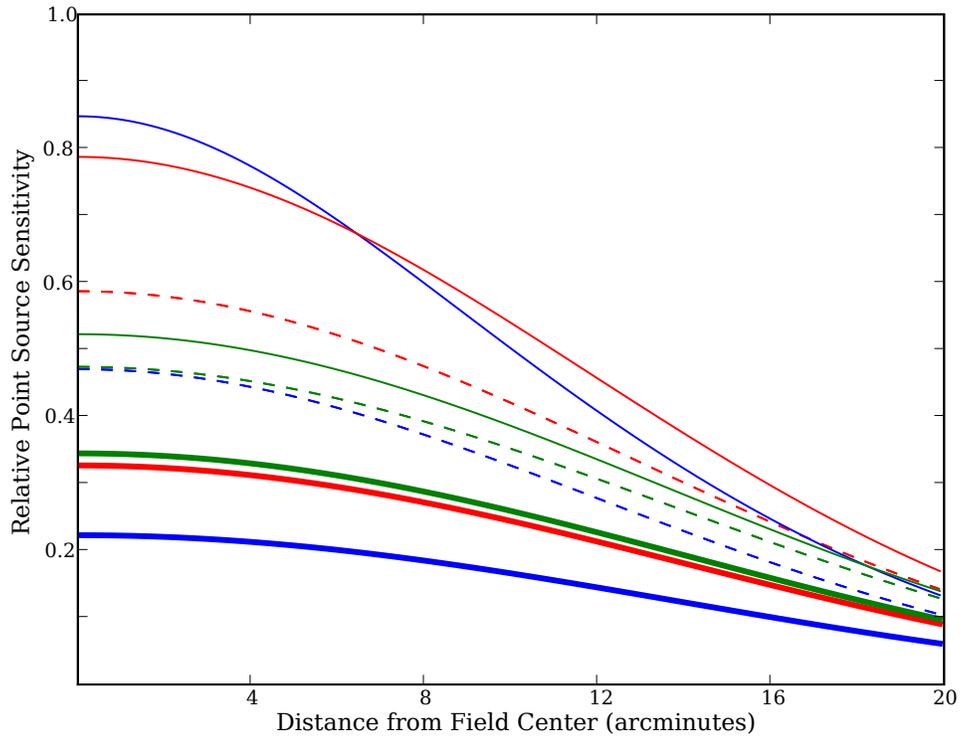}
\caption{Relative sensitivity for the three different image resolutions and 
three different source sizes: As in fig~\ref{rel20}  blue: full resolution 
($\sim 1.6$\arcsec),red: ($3.0$\arcsec), green ($6.0$\arcsec). The line solid lines are
for a source of 1\arcsec\ Gaussian FWHM; dash lines,a 3\arcsec\
FWHM and  thick solid lines, 7\arcsec\ FWHM. \label{size137}} 
\end{figure}
\clearpage

\begin{figure}
\plotone{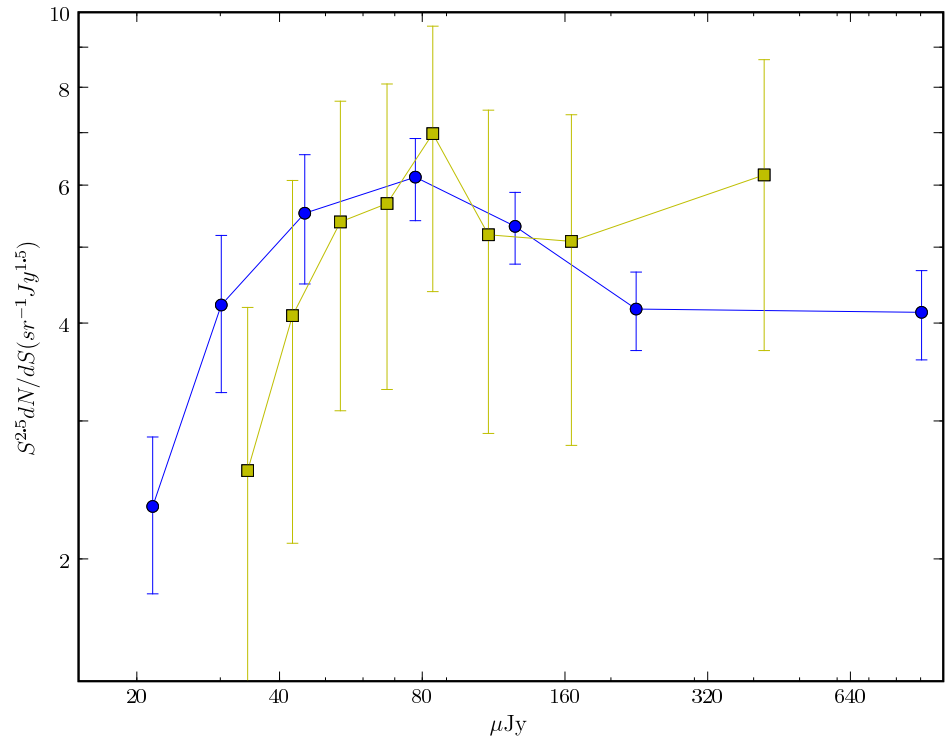}
\caption{The log N - Log S differential distribution uncorrected
for source resolution for our survey (blue dots) and the SA13
field (yellow squares). \label{sa13comp}}
\end{figure}
\clearpage

\begin{figure}
\plotone{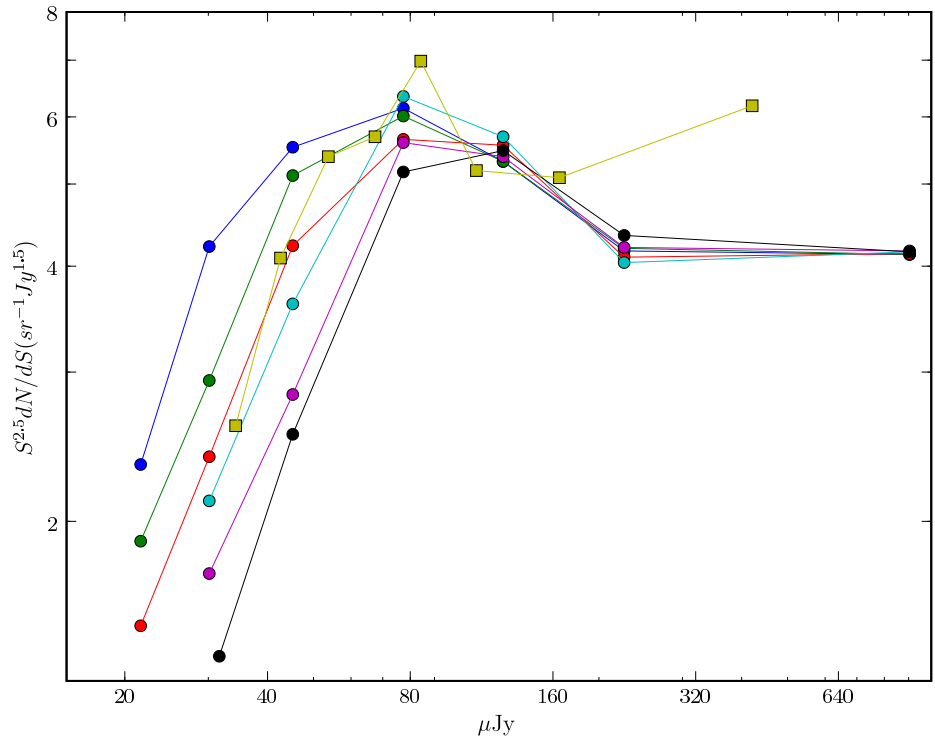}
\caption{The uncorrected log N - log S for our survey with various S/N
cutoffs (dots) and the SA13 results (yellow squares). For our
survey the curves are S/N= 5(blue), 6(green), 7(red), 8(cyan), 9(magenta) and
10(black). Note that all the curves turn over at low flux densities
but that the turnover shifts to higher flux densities with higher
S/N.  \label{rawlogNS}}
\end{figure}
\clearpage

\begin{figure}
\plotone{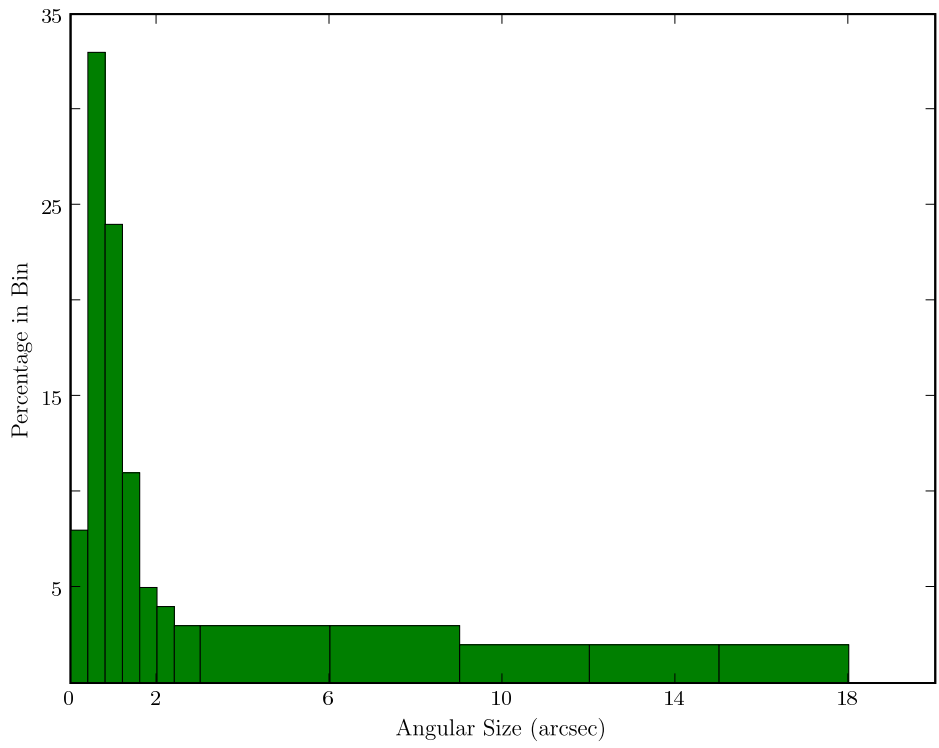}
\caption{The adopted angular size distribution used to correct the
log N - log S counts. Note that the percentages shown are for the
bin shown and are not normalized to a uniform bin size in arcseconds.
\label{angsize}}
\end{figure}
\clearpage

\begin{figure} 
\plotone{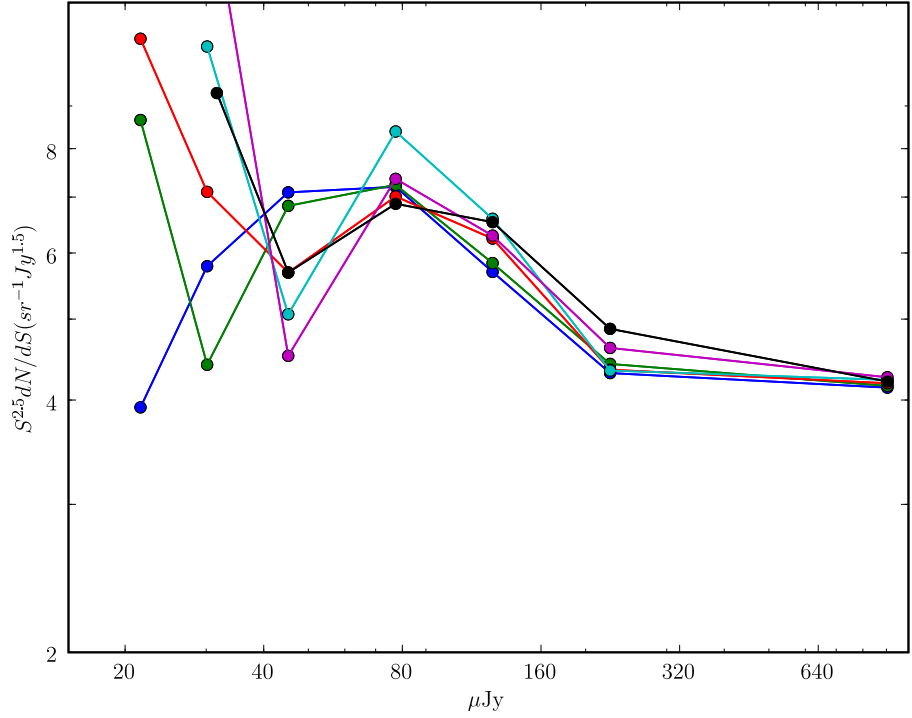}
\caption{The corrected log N - log S for our survey with various S/N
cutoffs (dots). The respective curves are S/N= 5(blue), 6(green),
7(red), 8(cyan), 
9(magenta) and 10(black). \label{corlogNS}}
\end{figure}
\clearpage

\begin{figure}
\plotone{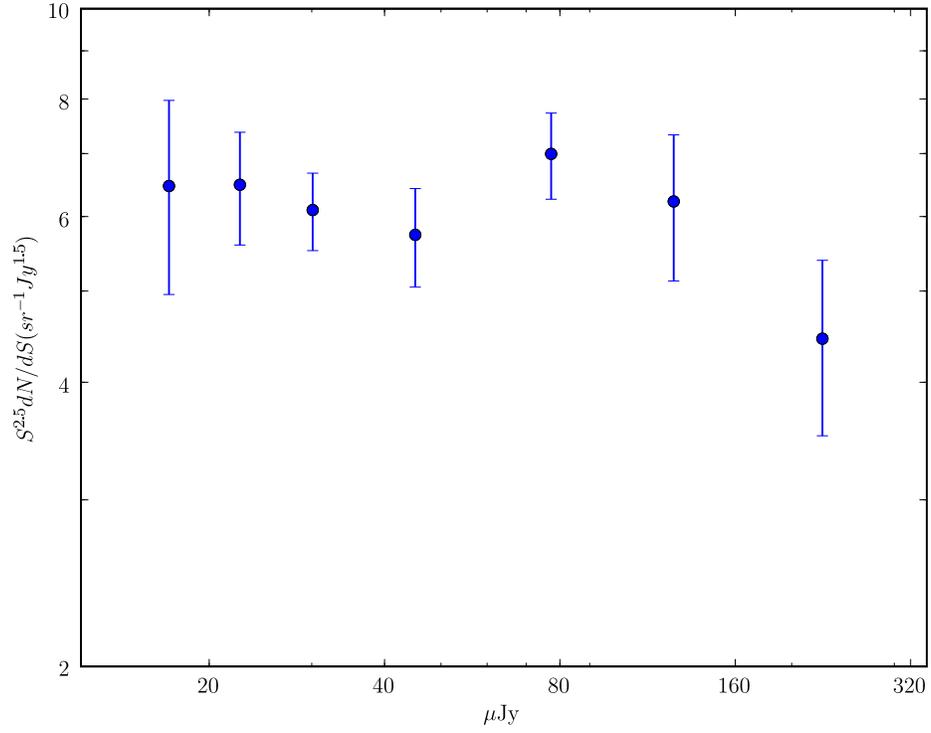}
\caption{Our final, corrected log N - log S using the
weighted sun of the estimates from our sliding cutoff window as 
described in the text. The error bars were determined from Monte
Carlo simulations. \label{final}}
\end{figure}
\clearpage

\begin{figure} 
\plotone{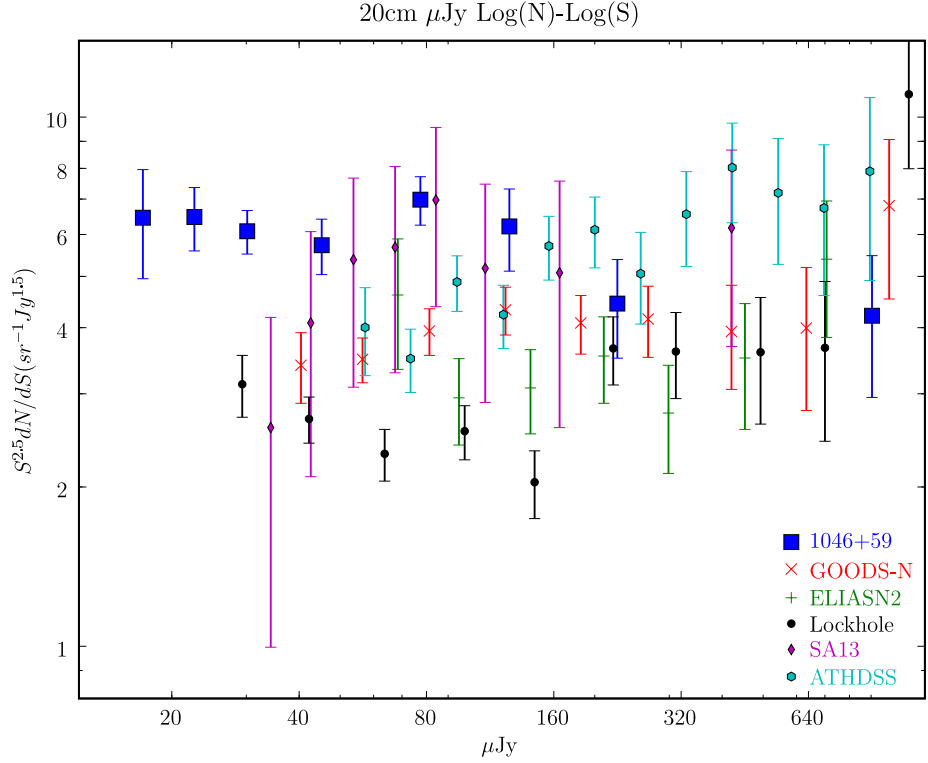}
\caption{
Recent log N - log S estimates for $S_{20cm} < 1$ mJy. The 1046+59
points are from this paper; the GOODS-N, ELIASN2, and Lockhole points are from
\citet{biggs}; SA13 data are from \citet{f06}; and ATHDSS data are from
\citet{huynh}.
\label{lognsall}}
\end{figure}
\clearpage

\end{document}